\documentclass[12pt]{article}
\begin{document}
\title{{\bf Does God So Love the Multiverse?}
\thanks{Alberta-Thy-20-07, arXiv:0801.0246, to be published in Melville
Y.~Stewart, ed., {\em Science and Religion in Dialogue} (Blackwell
Publishing Inc., Oxford), and in Melville Y.~Stewart and Fu Youde, eds.,
{\em Science and Religion: Current Dialogue} (Peking University Press,
Beijing, in Chinese), from a series of lectures sponsored by the
Templeton Foundation and given at Shandong University in Jinan, China,
autumn 2007; see also arXiv:0801.0245 and arXiv:0801.0247.}}
\author{
Don N. Page
\thanks{Internet address:
don@phys.ualberta.ca}
\\
Institute for Theoretical Physics\\
Department of Physics, University of Alberta\\
Room 238 CEB, 11322 -- 89 Avenue\\
Edmonton, Alberta, Canada T6G 2G7
}
\date{(2008 January 17)}

\maketitle
\large
\begin{abstract}
\baselineskip 17 pt

Monotheistic religions such as Judaism and Christianity affirm that God
loves all humans and created them in His image.  However, we have
learned from Darwin that we were not created separately from other life
on earth.  Some Christians opposed Darwinian evolution because it
undercut certain design arguments for the existence of God.  Today there
is the growing idea that the fine-tuned constants of physics might be
explained by a multiverse with very many different sets of constants of
physics.  Some Christians oppose the multiverse for similarly
undercutting other design arguments for the existence of God.  However,
undercutting one argument does not disprove its conclusion.  Here I
argue that multiverse ideas, though not automatically a solution to the
problems of physics, deserve serious consideration and are not in
conflict with Christian theology as I see it.

Although this paper as a whole is {\it addressed} primarily to
Christians in cosmology and others interested in the relation between
the multiverse and theism, it should be of {\it interest} to a wider
audience.  Proper subsets of this paper are addressed to other
Christians, to other theists, to other cosmologists, to other
scientists, and to others interested in the multiverse and theism.

\end{abstract}
\normalsize

\baselineskip 18.7 pt

\newpage

\section*{Preface Added to the Original}

\hspace{.20in} This paper was originally written as an expansion, for the
proceedings, of lectures given at Shandong University in Jinan, China, to
students and other academics who were interested in the relation between
science and religion.  Therefore, it employs a variety of philosophical
and theological argumentations that go beyond the usual scientific
reasoning of physics papers.  However, I believe that this paper may also
be of sufficient interest for many physicists to be included in the usual
physics eprint archives, despite the differences in style.  (Those who
are not interested in Christian theology may wish to skip the Bible
verses and stories, though they illustrate the claim of the breadth of
God's love. Similarly, those not interested in arithmetic may wish to
skip the example of the large finite set of pili-increasers and the offer
of a prize of perhaps \$1051 for the first member to be found of this set
of $10^{154}$ or so members, though this set illustrates how evidence for
a multiverse may be contained within a single universe.)

The written version of this paper has many different purposes.  One
primary purpose is to persuade my fellow Christians that multiverse ideas
are not necessarily opposed to theism and to Christianity.  A secondary
purpose, and one of the main reasons for posting it on the physics eprint
archives, is to show other scientists that not all Christians are opposed
to multiverse ideas, just as not all Christians are opposed to
evolutionary ideas, even though some Christians have opposed both.

I have long been open to multiverse ideas and remember asking Steven
Weinberg around 1983 what he thought of the idea that the constants of
physics might vary from one component of the quantum state of the
universe to another.  At that time, Weinberg said, ``I would find that
very depressing.''  However, Weinberg has come around to champion the
idea that the cosmological `constant' might vary across a
multiverse\footnote{Steven Weinberg, ``Anthropic Bound on the
Cosmological Constant,'' {\it Physical Review Letters} {\bf 59},
2607-2610 (1987); Hugo Martel, Paul R.~Shapiro, and Steven Weinberg,
``Likely Values of the Cosmological Constant,'' {\it The Astrophysical
Journal} {\bf 492}, 29-40, astro-ph/9701099,
$<$http://arxiv.org/abs/astro-ph/9701099$>$.}.

Even more recently, Weinberg has made the following
argument\footnote{Steven Weinberg, ``Living in the Multiverse,'' in
Bernard Carr, ed., {\em Universe or Multiverse?} (Cambridge University
Press, Cambridge, 2007), pp. 29-42, hep-th/0511037,
$<$http://arxiv.org/abs/hep-th/0511037$>$.}:
\begin{quotation}
``Finally, I have heard the objection that, in trying to explain why the
laws of nature are so well suited for the appearance and evolution of
life, anthropic arguments take on some of the flavor of religion.  I
think that just the opposite is the case.  Just as Darwin and Wallace 
explained how the wonderful adaptations of living forms could arise
without supernatural intervention, so the string landscape may explain
how the constants of nature that we observe can take values suitable for
life without being fine-tuned by a benevolent creator.  I found this
parallel well understood in a surprising place, a New York Times op-ed
article by Christoph Sch\"{o}nborn, Cardinal Archbishop of
Vienna.\footnote{C.~Sch\"{o}nborn, N.~Y.~Times, 7 July 2005, p.~A23.} 
His article concludes as follows:

\begin{quotation}
`Now, at the beginning of the 21st century, faced with scientific claims
like neo-Darwinism and the multiverse hypothesis in cosmology invented to
avoid the overwhelming evidence for purpose and design found in modern
science, the Catholic Church will again defend human nature by
proclaiming that the immanent design evident in nature is real. 
Scientific theories that try to explain away the appearance of design as
the result of `chance and necessity' are not scientific at all, but, as
John Paul put it, an abdication of human intelligence.'
\end{quotation}

``It's nice to see work in cosmology get some of the attention given these
days to evolution, but of course it is not religious preconceptions like
these that can decide any issues in science.''
\end{quotation}

Here I wish to go on record as a Christian who respectfully differs from
Cardinal Sch\"{o}nborn's opinion that the multiverse idea was ``invented
to avoid the overwhelming evidence for purpose and design found in modern
science'' and that it is ``an abdication of human intelligence.'' 
Different multiverse theories were invented for different reasons, and
there are intelligent reasons for investigating them.  Furthermore, while
I agree with Weinberg that multiverse theories would not require that the
`constants' of nature be fine tuned, I would also insist that they do not
preclude the possibility that the entire multiverse was designed at a
higher level, say by a benevolent Creator.

\newpage

\section{God's Love for All Humans}

\hspace{.20in} A central point of Judaism and Christianity is that God
loves everyone.  For Christians, one of the most famous verses in the
Bible is in the Gospel of John, John 3:16:  ``For God so loved the world
that He gave His only begotten Son, that whoever believes in Him should
not perish but have everlasting life'' \cite{Bible}.  However, this idea
starts way back in the Old Testament (the first part of the Bible, the
part accepted by both Jews and Christians).

In Genesis, the first book of the Old Testament, God began a revelation
through the family of Abraham, Isaac, and Jacob (who became the
Israelites, Hebrews, and/or Jews).  In God's original call to Abraham in
Genesis 12:1-3, He said, ``I will make you a great nation; I will bless
you and make your name great; \ldots and in you all the families of the
earth shall be blessed.''  As a personal example of this, my family is
not Jewish, and yet academically I have been marvelously blessed by the
remarkable contributions of a vast number of brilliant Jewish
scientists, and spiritually I have been immeasurably blessed by the
teachings, life, death, and resurrection of the Jewish rabbi Yeshua,
known in English as Jesus, who we Christians believe to be the Christ or
Messiah, the anointed Son of God and Savior.

Later in the Old Testament, in the book of Ruth, God's love was extended
beyond the Israelites to Ruth, a woman of the foreign country of Moab who
married an Israelite who had settled there during a famine in Israel. 
After both Ruth and her mother-in-law Naomi were widowed, Ruth moved with
Naomi to Israel and became an ancestor to Israel's greatest king,
David, as well as an ancestor of Jesus.  

Jonah is another book in the Old Testament stressing the extension of
God's love beyond the Israelites.  This famous short story tells of the
prophet Jonah who was sent by God to preach against the evils of Nineveh,
a great city of ancient Assyria, on the eastern bank of the Tigris River
at the location of the present city of Mosul, Iraq.  As an Israelite
enemy of Nineveh, Jonah wanted Nineveh to remain unrepentant and be
destroyed by God.  Therefore, initially he refused to go warn that evil
city but rather fled in the opposite direction.  As Jonah later admitted
to God, ``Therefore I fled previously to Tarshish; for I know that You
are a gracious and merciful God, slow to anger and abundant in
lovingkindness, One who relents from doing harm.''  However, through a
remarkable event God got Jonah to obey Him, and the Ninevites repented
and were spared, still to the bitter anger of Jonah.  The book closes
with God's emphasizing His love for the foreign city:  ``And should I not
pity Nineveh, that great city, in which are more than one hundred and
twenty thousand persons who cannot discern between their right hand and
their left---and much livestock?''

In the New Testament, God sent His Son Jesus Christ to live, die on the
cross for our sins, and defeat death by His resurrection, to bring
forgiveness and salvation to all of us who cannot meet God's standards as
set down by the laws God had given to the Israelites in the Old
Testament.  As the Apostle Paul expressed it in Romans 3:23-24:  ``For
all have sinned and fall short of the glory of God, being justified
freely by His grace through the redemption that is in Christ Jesus.'' 
This Gospel message of salvation by faith was offered to all people, not
just the Jews.  The last recorded words of Jesus Christ, in the Gospel of
Matthew 28:18-20, are these:  ``All authority has been given to Me in
heaven and on earth.  Go therefore and make disciples of all the nations,
baptizing them in the name of the Father and of the Son and of the Holy
Spirit, teaching them to observe all things that I have commanded you;
and lo, I am with you always, even to the end of the age.''

The first book of the history of Christianity, after the Gospel stories
of the life, death, and resurrection of Jesus, is the Acts of the
Apostles.  It particularly records how the Apostles Peter and Paul began
the work of extending the Gospel message beyond the Israelites to all
nations.

God not only loves all people, but He also said that He created all
people in His own image.  As it is written in the first book of the
Bible, in Genesis 1:27, ``So God created man in His own image; in the
image of God He created him; male and female He created them.''

The question arises as to how unique does that make us.  The Bible
certainly emphasizes that the image of God extends to all humans.  But
are we created entirely separately from the rest of creation?

Some have taken the image of God for humans to imply that God created us
individually and separately from other living beings.  However, Darwin's
theory of evolution suggests that we are related to the rest of life.  It
also suggests that we humans were not separately created by an individual
act, independent of the creation of the remainder of the earth's
biosphere.

\section{Parallels Between Evolution and Multiverse Ideas}

\hspace{.20in} When Darwin proposed evolution, many conservative
Christians accepted it.  One famous example was Benjamin B. Warfield
(1851-1921), the conservative Christian theologian and principal of
Princeton Seminary from 1887 to 1921.  Warfield wrote the chapter on
``The Deity of Christ'' in {\em The Fundamentals}, from which the term
Fundamentalism arose.  Thus one of the most famous original
Fundamentalists accepted Darwinian evolution, writing \cite{Warfield},
``I am free to say, for myself, that I do not think that there is any
general statement in the Bible or any part of the account of creation,
either as given in Genesis 1 and 2 or elsewhere alluded to, that need be
opposed to evolution.''

However, many Christians later came to oppose evolution, perhaps most
famously some other Fundamentalists.  Although there were many reasons
for this, which I cannot get into here, one possible reason is that
evolution did remove one particular design argument for the existence of
God, that all of the marvelously many different species of living things
on earth had been separately designed and created by God.  Nevertheless,
evolution did not disprove the existence of God or of some overall
design.  Indeed, there are many leading theologians and scientists today
that accept both evolution and creation by God, such as Francis Collins,
the head of the Human Genome Project \cite{FCollins}.

It seems to me that there may be a parallel development occurring
today.  Before Darwin, some Christians took the marvels of humanity as
evidence of separate and individual design.  Now, some Christians take
the marvels of the fine-tuning of the constants of physics as evidence
of theism and often of separate and individual design of these constants
by God \cite{Ross83,Ross88,Ross89,Swinburne90,Swinburne91,Ross01,Manson,
Holder,Mann,FCollins}.  Here I wish to argue that this could be equally
mistaken.

I have found that my views are rather similar to those of a minority of
theists, notably John Leslie \cite{Leslie1,Leslie2}, Stephen Barr
\cite{Barr}, Robin Collins \cite{Collins01,Collins05,Collins07}, Gerald
Cleaver \cite{Cleaver03,Cleaver06}, Klaas Kraay \cite{Strand-Kraay}, and
others that Kraay cites, who break tradition and argue that a multiverse
could reveal an even more grand design of the universe, since the
physical process that generates the multiverse would have to have
suitable basic laws and initial conditions to produce any life at all
(no matter what the constants of physics are, since often they seem to
be fine tuned for several different reasons \cite{Leslie1,Leslie2}). The
laws and initial conditions would apparently have to be even more
special to produce not just life, but life like ours observing the order
we actually do see around us.  Leslie, Barr, Collins, Cleaver, Kraay,
and others claim that since God is infinitely creative, it makes sense
to say that He might create a physical reality much larger than the
single visible part of the universe or multiverse that we can observe
directly.

\section{Fine Tuning in Our Universe}

\hspace{.20in} Now it does seem to be true that we could not be here if
many of the constants of physics were significantly different, so that
in our part of the universe, the constants of physics do in fact seem to
be fine tuned for our kind of life.  This is generally agreed upon both
by those who attempt to use this fine tuning to support theism, as in
the references above, and by many scientists who are usually neutral or
opposed to such an attempt \cite{Carter74,Carr-Rees,Davies82,Carter83,
BarTip,Davies92,Rees,Barrow,Bostrom,Susskind,Davies07,Carr,Carter07a,
Carter07b}.  Of course no one knows what other forms of life might be
possible if the constants of physics were significantly different, but
the general consensus seems to be that if would be very difficult to
imagine the possibility of any complex life at all existing if certain
combinations of the constants of physics were greatly different.

For example, one of the most remarkable fine tunings is the value of the
cosmological constant or energy density of the `dark energy' responsible
for the current acceleration of distant galaxies away from each other. 
(Even though a positive cosmological constant corresponds to positive
energy density which normally would gravitate or tend to pull things
together, it is also accompanied by tension or negative pressure that
antigravitates by an amount that is larger by a factor of three, the
number of dimensions of space.  Thus the net observed effect of the
positive cosmological constant or dark energy is antigravity, the
gravitational repulsion of otherwise empty space that causes distant
galaxies now to be accelerating apart.)

Measurements show that the cosmological constant is more than 120 orders
of magnitude smaller than unity in certain natural units (called Planck
units, obtained by setting to unity the speed of light, Planck's quantum
constant of action, and Newton's gravitational constant).  With the other
constants kept fixed, it would be difficult to have a universe with
gravitationally formed structures lasting long enough for life if the
cosmological constant were even just a few orders of magnitude (powers of
10) larger than its observed value.  But even if one tuned the other
constants to allow the possibility of such structures when the
cosmological constant has a value many orders of magnitude larger than
its observed value, one still seems to need it to be many orders smaller
than unity.  So one does not see how to avoid at least some significant
amount of fine tuning of this parameter.  (Basically, if the cosmological
constant were of the order of unity in the natural Planck units, the
spacetime of the universe would always have large quantum mechanical
fluctuations, and no one knows any plausible way to have persisting
complex structures that one could call life in such a case.)

Another constant that is many orders of magnitude away from unity, in
this case about 36 orders of magnitude larger than unity, is the ratio of
the electrostatic repulsion to the gravitational attraction between two
protons (the nuclei at the centers of hydrogen atoms).  With other
constants kept fixed, it seems that one could not have the types of stars
that appear to be necessary for life if this constant differed by much
more than even one order of magnitude (factor of 10) from its actual
value.  Again one could try to imagine a universe hospitable to some
other form of life when this constant is significantly different by also
tuning other constants to an appropriate range, but again it seems that
complex life of any form relying mainly on the electromagnetic and
gravitational forces would be impossible if this constant were close to
unity.  (Then it seems that one could not have stars, planets, and living
organisms with large numbers of atoms, since the number of atoms in such
structures generally scales as a positive power of this constant and
would approach some small number near unity if this constant were itself
near unity \cite{Carter07a,Carter07b}.)

Martin Rees \cite{Rees} discusses in much more detail these two constants
and four others in our universe that are crucial for its properties. 
Life as we know it would apparently be impossible if any one of them were
greatly different (with the others held fixed).  So although it might not
be necessary for all of them to have their observed values, there are
some combinations of them that apparently could not be very much
different and yet give a universe with life, at least life at all similar
to present life on earth.

\section{Explanations for Fine Tuning}

\hspace{.20in} So there is a general consensus that there is at least
some fine tuning of the constants of physics in our part of the
universe, though not that all of the constants had to have values close
to what we observe.  But what is the explanation for this phenomenon? 
There are three general types of explanations that are often put
forward.

Some suggest that the fine tuning was done by a separate act of God to
allow life.  Others say that it is presumably an accidental fluke.  And
yet others propose that it arises from a huge multiverse of very many
different possible constants of physics.  It is also noted in several of
the references that I have given, such as \cite{Leslie1,Page-Carr}, that
the three explanations are not mutually exclusive, so that virtually any
combination of them is logically possible.  However, it is the
multiverse explanation that is now rapidly growing in favor, though not
without a lot of opposition from both theists and nontheists.

One must quickly point out that each of these three explanations really
stands for a class of explanations, so that one should actually compare
specific proposals taken from these classes rather than the classes
themselves.  Theists of different theological convictions might propose
different ideas of how God would choose the constants.  Those saying that
the fine tuning is a fluke might say that the constants are determined by
any number of different mathematical structures that just happened to
give biophilic values, or they might propose that there is truly some
random process determining the constants in some way not derivable from
any simple mathematical structure.  And of course there are a huge number
of possible multiverse theories.

Some multiverse theories seem to me to be too general to be plausible,
such as the idea of David Lewis \cite{Lewis} that all logical
possibilities actually exist, or the original idea of Max Tegmark
\cite{Tegmark} that all mathematical structures have physical reality. 
These would seem to leave it unexplained why what we see has the order
that it does, whereas a random possibility from all logical possibilities
or from all mathematical structures would surely be far more chaotic
\cite{Leslie1,Leslie2,Barr,Page-Carr} (though I now admit that my
objection to Tegmark's original Level IV multiverse as logically
inconsistent was wrong, since I misinterpreted each mathematical
structure to be a description of reality rather than being simply a part
of reality).  However, there might be other multiverse theories that are
more explanatory of the order that we do observe, perhaps arising
naturally out of elegant but specific laws of nature.  For example,
Tegmark's more recent ideas \cite{Teg07} that physical reality might be
restricted to the mathematical structures of computable functions seems
to me more hopeful, though I would disagree with his view that physical
reality is just mathematical structure or syntax and instead believe that
it includes at least the semantics of consciousness.

One natural way to get a multiverse is to have a universe so large that
highly varied conditions occur somewhere.  Another is from Everett
many-worlds \cite{many-worlds}, that all the quantum possibilities are
actually realized.  However, those possibilities do not necessarily give
varying constants of physics.

One scenario that seems more hopeful is to get multiverses from inflation
\cite{Linde,Guth,Vilenkin}, which is a very rapid exponential expansion
of the early universe that may make the universe enormously larger than
what we can observe of it.  If the inflationary scenario can include
phase transitions, and if the constants of physics can differ across
phase transitions, inflation tends to produce all such possibilities.

Recently it has been realized that string/M theory apparently leads to a
huge multiverse of $10^{500}$ or so different vacua or sets of constants
\cite{DouKac}.  This would apparently be enough for the constants we see
to occur somewhere (maybe once per $10^{200}$ vacua or so).  Then perhaps
$10^{300}$ or so vacua would fit what we see.

If only one universe in $10^M$ could fit our observations, but if $10^N$
different universes exist in the multiverse, then it might not be
surprising that what we observe exists if $N > M$.  E.g., in the
previous paragraph, I was saying that perhaps $N$ is around 500 and $M$
is around 200, so then indeed $N > M$.  However, the actual numbers are
known very poorly \cite{DouKac}.  We really don't yet know whether $N >
M$ in string/M theory, but that seems plausible.  Then what we see could
be explained without its having to be individually selected.

One might still ask whether the multiverse explanation always works,
assuming that it has enough universes (e.g., $N > M$).  Is it sufficient
to explain what we see by a multiverse theory in which there are enough
different conditions that ours necessarily occurs somewhere?

I would say no, but rather that there is the further requirement that the
conditions we observe should not be too rare out of all the conditions
that are observed over the entire multiverse.  A theory making our
observations too rare should not be considered a good theory.

Good theories should be both intrinsically plausible and fit
observations.  Intrinsic plausibility is quantified by what is called
the {\it a priori} probability of the theory, the probability that one
might assign to it from purely theoretical background knowledge, without
considering any observations.  The fit to observations is quantified by
the conditional probability of the observation given the theory, what is
called the {\it likelihood}.  Then the probability of the theory after
taking into consideration the observation, what is called the {\it a
posteriori} probability of the theory, is given by Bayes' theorem as
being proportional to the product of the {\it a priori} probability and
the likelihood \cite{Swinburne}.

I take the {\it a priori} probabilities of theories (intrinsic
plausibilities before considering the observations) to be subjective but
to be generally assigned higher values for simpler theories, by the
principle that is called Occam's razor.  One problem with this is that
David Deutsch \cite{Deutsch,Deutsch1} notes that simplicity depends on
one's background knowledge that itself depends on the laws of physics.

The {\it likelihood} of a theory is itself neither the {\it a priori} or
the {\it a posteriori} probability of the theory, but rather the
conditional probability, not of the theory, but of the observation given
the theory.  A theory that uniquely gives one's observation would have
unit likelihood but might have very low {\it a priori} probability.

For example, consider an extreme solipsistic theory that only one's
actual momentary observation exists, not anyone's else's or even any of
one's own in either the past or the future, and perhaps not even that an
external world exists at all.  This theory would predict that
observation with certainty if it were correct.  (If the theory were
true, certainly the observation would be that single one predicted by
the theory.)  Therefore, for that observation the likelihood is unity. 
However, such an extreme solipsistic theory, giving all the details of
one's observation or conscious perception without an external world
giving other observations, would surely be highly complex and so would
be viewed as extremely implausible, much more implausible than an
alternate theory in which the observation resulted from the existence of
an external world that also gives other observations.  Therefore, this
extreme solipsistic theory would be assigned very low {\it a priori}
probability.

At the other extreme, consider the simple theory that predicts all
possible observations equally (arguably a consequence of something like
the modal realism of David Lewis \cite{Lewis}).  Since this theory is so
simple, it might be assigned a high {\it a priori} probability, but then
because of the enormous number of observations it predicts with equal
probability, it would give very low likelihood.

\section{Applying Bayes' Theorem}

\hspace{.20in} Let us consider a simple example with three possible
theories and use Bayes' theorem to calculate the resulting {\it a
posteriori} probabilities.  Suppose that we have theory $T_1$ with {\it a
priori} probability 0.000\,000\,1 that would give probability 1 for what
we see (unit likelihood; the observation would be certain {\it if} this
theory were true), theory $T_2$ with {\it a priori} probability 0.001
that would give probability 0.01 for what we see (1\% likelihood), and
theory $T_3$ with {\it a priori} probability 0.998\,999\,9 that would
give probability 0.000\,000\,1 for what we see (0.000\,01\% likelihood). 
Assume for simplicity that these three theories exhaust all possible
theories.

Then the product of the {\it a priori} probability and of the likelihood
for the first theory, $T_1$, is 0.000\,000\,1, for $T_2$ is 0.000\,01,
and for $T_3$ is nearly 0.000\,000\,1.  The {\it a posteriori}
probabilities of these three theories, by Bayes' theorem, is
proportional to these products, so all we need to do is to normalize the
products by dividing each of them by their sum, which is nearly
0.000\,010\,2.  After dividing by this, we then get that to two-digit
accuracy, the approximate {\it a posteriori} probability of $T_1$ is
0.01, of $T_2$ is 0.98, and of $T_3$ is 0.01.  That is, after the
observation, we would think that theories $T_1$ and $T_3$ each have only
about 1\% probability of being correct, whereas in the end theory $T_2$
is seen to be 98\% probable.

In this case, neither $T_1$ with its unit likelihood, nor $T_3$ with its
nearly unit {\it a priori} probability, gains the highest {\it a
posteriori} probability, which instead goes to the compromise theory
$T_2$, which had both its {\it a priori} probability (0.001) and its
likelihood (0.01) rather low.  Of course, there is no guarantee that a
compromise theory will gain the greatest {\it a posteriori} probability,
since the result depends on the particular {\it a priori} probabilities
and likelihoods of the theories being tested.  However, if the theories
were ranked in order from most specifically predictive, say giving unit
likelihood or unit probability for the specific observation, to the
simplest theory with the highest {\it a priori} probability, it would
seem unusual for either of the two extreme theories to end up with the
greatest {\it a posteriori} probability.  In this way I would be
surprised if the theory for the universe or multiverse with the greatest
{\it a posteriori} probability turned out to be either the most
specifically predictive theory (giving precisely one's own observation
with certainty, and no probability of any other possible observation one
might have had instead) or the simplest (say predicting all possible
observations with equal likelihood).

Suppose that as a theist one wants to assign {\it a priori}
probabilities to different theories of what kind of universe or
multiverse God might want to create.  One might guess that God might
want to create the simplest or most elegant universe or multiverse that
has beings like us that He considers to be in His image or that can have
fellowship with Him or that has some other moral good in it.  Then the
theist might try to guess whether it would be simpler for God to create
just one universe with a single set of constants of physics, or whether
it would be simpler for Him to create a multiverse with a variety of
sets of constants.

If there were a simple principle for choosing the particular constants
that we observe (e.g., simple laws of physics that give these uniquely),
then one might suppose that God would prefer just using those laws and
creating a unique set of constants.  However, we don't see any reason why
the constants we observe should be uniquely preferred over other
possibilities.  Of course, this could easily just be a failure of our
imagination and knowledge, so we might continue looking.  Even
nontheistic scientists like David Gross might prefer a single-universe
theory with a single set of constants of physics and persist in looking
for simple laws of physics that would give this unique result, following
Winston Churchill in saying, ``Never, never, never, never, never give
up'' \cite{Woit}.

On the other hand, not finding a simple principle from which one can
deduce uniquely the constants of physics we observe, and having some
potential theories like string/M theory that strongly suggest a
multiverse instead, might lead many, both theists and nontheists, to
look for a multiverse theory instead.  From a theistic perspective, it
might seem simpler (or better in other regards) for God to choose a
variety of the sets of constants of physics, a multiverse rather than a
single universe.

The situation is reminiscent of that facing Johannes Kepler, who
attempted to explain both the number of planets and their orbits.  He was
remarkably successful in describing the shape of the orbits, but his
attempted explanation that the number was six failed, as was directly
shown by the discovery of more planets.  Now we know that different solar
systems can have different numbers of planets, so there would not have
been a way for Kepler successfully to explain a unique value.  We are
faced with a similar situation with the constants of physics, in that at
present we do not know whether they are like the orbital shapes, which
can be explained by simple principles (e.g., Newton's law of universal
gravitation) or whether they are like the number of planets, which can
have different values for different systems.  I do not see that it
ultimately makes a significant difference theologically which way the
answer turns out, but as seekers of the truth we would like to find what
is actually the case, or in theistic terms, what God actually did.  Given
our present knowledge, to me it currently seems simpler to hypothesize
that God created a multiverse, and I would argue that that is a
theologically acceptable option for Christians and other theists to
consider.

\section{Toy Multiverse Model from Arithmetic}

\hspace{.20in} For those of you who are mathematically inclined, the
following example from arithmetic might be a helpful analogue,
illustrating how observations within just one universe can give evidence
of whether or not other universes also exist.

Suppose that we imagine possible universes to be analogous to positive
integers, and one's observed universe to be analogous to a particular
integer.  For example, suppose that there were a simple prescription for
translating the constants of physics into a single integer (e.g., some
binary string encoding the values of the constants).  Then we might seek
the simplest explanation for the integer corresponding to the observed
universe or to the observed constants of physics in that universe.

If the integer itself were simple, such as $2^{2^{2^{2^2}}}$, which is
very large (19\,729 decimal digits) but quite simple to state, one might
suppose that the simplest explanation would be an explanation that
postulates just the existence of one universe and hence just that one
simple integer.  On the other hand, if the integer were a very complex
member of a simple set of integers, then it might be simpler to
postulate that the entire set exists.

For example, any large positive integer, no matter how complex, is a
member of the set of all positive integers, which is a very simple set. 
If the observed universe corresponded to such a very complex integer (not
given by a simple algorithm such as the example $2^{2^{2^{2^2}}}$ above),
one might suppose that it would be simpler to suppose that the much
simpler set of all possible positive integers existed.  (Note that
although this set is infinite, it does not encompass negative integers,
rational numbers, algebraic numbers, real numbers, complex numbers,
quaternions, or any number of other mathematical structures, so that even
an infinite multiverse need not encompass all logical possibilities.)

However, one might worry that if the set of possibilities is infinite
(even the highly limited infinity of the positive integers), then the
probability of getting the particular observed integer would be zero, so
the hypothesis or theory that the multiverse corresponded to the simple
set of all possible positive integers would be assigned zero likelihood. 
(One way to avoid this would be to put a simple normalizable weight upon
all the positive integers, so that the total weight is finite, allowing
one to get nonzero probabilities for all finite positive integers, but
for now let us try something else.)

To allow equal weights for all positive integers corresponding to
universes in a multiverse, one might hypothesize that the multiverse
corresponds to some finite set of positive integers.  As an example of a
large but finite simple set, consider the following set of what I call
{\it pili-increasers:  positive integers $n$ for which $\pi(n$)/li($n$)
is both greater than one and greater than that of any smaller $n$.} 
Here $\pi(n$) is the number of primes not greater than $n$, and li($n$)
is the Cauchy principal value of the integral of the reciprocal of the
natural logarithm of $x$ from 0 to $n$, which by the prime number
theorem gives a good asymptotic estimate for $\pi(n$) but which
oscillates around $\pi(n$) an infinite number of times as $n$ increases
\cite{Littlewood}.  

This set of pili-increasers is a fairly simple set to define, and yet it
apparently has more members than particles in the observable universe
(the part of our universe we can see from light emitted after most of
the electrons and ions combined in the early universe to make it
transparent, which contains roughly $10^{90}$ particles).  Each member
of this set of pili-increasers likely has more than a thousand binary
digits, so almost all individual members of this set are almost
certainly much more complex than the entire set.  (Imagine trying to
memorize one of these thousand-binary-digit numbers, in comparison with
trying to memorize my short definition of the set.)

The smallest member of this set of pili-increasers has been estimated to
be about $1.398\times 10^{316}$ \cite{Bays-Hudson}.  Smaller candidate
pili-increasers, for example near $10^{190}$, have not yet been
rigorously ruled out, although such smaller numbers appear unlikely
actually to be pili-increasers.  (Ref.\ \cite{Bays-Hudson} only gives
rigorous upper bounds on the smallest pili-increaser, the smallest
$x\geq 2$ for which $\pi(x)>$li($x$).)

Less is known about the largest pili-increaser, which I call the {\it
pili-maximizer}, the positive integer $n$ that maximizes the ratio of
$\pi(n$) to li($n$).  However, from Figure 2b of \cite{Bays-Hudson}, it
appears that it might be about $10^{311}$ larger than the most likely
candidate for the smallest pili-increaser (i.e., larger by roughly one
part in 140\,000 of the smallest pili-increaser that is likely to be
near $1.4\times 10^{316}$), and give a maximum value of $\pi(n$)/li($n$)
approximately $1+10^{-154}$.  The number of pili-increasers would be
expected to be about twice the value of $\pi(n$)$-$li($n$) at the
pili-maximizer, which from Figure 2b of \cite{Bays-Hudson} appears to be
nearly $10^{154}$.  Thus one might expect somewhat more than $10^{154}$
pili-maximizers, but spread over a range of roughly $10^{311}$ from the
smallest to the largest, so that if one picked a number at random within
that range, the probability might be of the order of $10^{-157}$, or one
part in more than thirteen trillions multiplied together, of picking a
pili-increaser.

In fact, no one knows the precise value of any of the individual members
of this set of pili-increasers.  I am hereby offering to pay \$1 U.S.\
for each of the binary digits (probably 1051, if the estimates above are
indeed applicable for all the members and are not just upper bounds for
the smallest member) of the first member of this set that is found
explicitly as a binary integer and is proved to be a member of the set. 
I am also offering a more attainable prize of \$1 U.S.\ for each binary
digit proved to be correct for the logarithm to base 2 of the
pili-maximizer (up to as many binary digits of this logarithm as are
needed to give the integer precisely, perhaps 1061; I won't pay any more
for the arbitrarily large number of additional digits that can be found
trivially if and when the exact pili-maximizer integer is found).

Although there is thus apparently an enormous number of members of this
simple finite set, they are also apparently extremely rare among all
integers in the same range.  If one chose a random integer between the
smallest and the largest member of this set, the chance that it would be
in the set would apparently be smaller than the chance of choosing one
particular particle out of all those in the observable universe. 
Therefore, if in some hypothetical universe some observer found the
constants of the physics translated into one member of that extremely
sparse set, it would certainly be strong evidence of fine tuning, strong
evidence against a multiverse corresponding to all possible positive
integers and even against a finite multiverse corresponding to all
positive integers within one part in 140\,000 of the smallest
pili-increaser.

If one then further found that this integer were a special member of the
set of pili-increasers, say the smallest or the largest, that would be
such extreme fine tuning that it would be strong evidence for the
hypothesis of a single value or a single universe (or at least for just a
small number of integer values or universes).

On the other hand, if the evidence were strong that this number were not
a special member of the set (though there is no algorithm for proving
this), then that particular integer would apparently be much more complex
than the simple set of all pili-increasers.  In this case, I would say
that the observer would be justified to postulate the existence of a
multiverse giving the entire simple set, and not just the particular
complex though fine-tuned example directly observed.

Although one might accuse this observer of being extravagant in
postulating this enormous set of perhaps roughly $10^{154}$ multiverses,
in reality the entire set would be much simpler than the individual
member observed (at least if the individual member were in fact a random
member of the set, as almost all members are, though there is no
algorithm for proving that any particular member is).  This is one sense
in which an observation in one single universe can be more simply
explained by postulating a large set of universes in a multiverse,
though without needing to postulate that all possible universes exist.

This simple toy example from arithmetic illustrates the fact that in
principle even from one single observation result (the one integer) in a
single universe, one can gain strong evidence (though even here not a
rigorous proof) for the much greater simplicity of either a
single-universe or a multiverse hypothesis, depending on what is
observed.  Of course, in practice, even in this toy example it would be
very difficult to calculate whether the integer were in the simple set
of pili-increasers and then whether it were a special member such as the
smallest or the largest.  However, one might develop a suspicion that it
might be a member if one found that the integer were a prime (as all
pili-increasers are) in the relatively narrow range between the smallest
member and the largest member, for which one can imagine fairly good
approximate estimates being made in the foreseeable future.  Then if one
found that it were within a narrow uncertainty of the estimate for one
of the endpoints of the range, one might suspect that it is at the
endpoint and interpret that as evidence for the single-universe
hypothesis, whereas if one found that it were not near an endpoint, one
might take that as evidence for the multiverse hypothesis that the
entire simple set existed.

Thus one can gain theoretical evidence for (or against) a multiverse
even from observations restricted to one single universe, and in
principle one might even gain evidence of how large the multiverse is. 
One's limited observations within a single universe are just a small
sample of the whole, but if it is a sufficiently rich sample, it can
give much information about the rest that is not directly observed.

\section{The Growth of Our Knowledge of the Universe}

\hspace{.20in} Our whole growth of knowledge of the universe has been an
expansion of its scope.  As one grows as an infant, one rapidly grows
beyond the view that one's present observation is all that is real, as
one develops memories about the past and anticipations of the future. 
One then goes beyond solipsism and gains an understanding that other
persons or observers exist as well.  In the early stage of human
development, there was the focus on one's family, which was then
gradually extended to one's tribe, one's nation, one's race, and, one
might hope, to all humans.  But then when one further considers what
other conscious observations may be going on, one might well believe
that consciousness extends to other creatures, such as other animals.

Of course, one's direct observation never extends beyond one's own
immediate conscious perception, so one can never prove that there are
past or future perceptions as well (and I know philosophers who do not
believe the future exists).  Similarly, one can never directly
experience even the present conscious perceptions of another, which
engenders the problem of other minds in philosophy.  Nevertheless, most
of us believe that we have fairly good indirect evidence for the
existence of other conscious experiences, at least for other humans on
earth with whom we can communicate, though it is logically possible that
neither they nor any external world actually exists.  (For me, I believe
that it is much simpler to explain the details of my present observation
or conscious perception or experience by assuming that an external world
and other conscious experiences also exist, than by assuming that just
my own momentary conscious perception exists.)

We may now extend the reasoning to suppose that if the universe is large
enough, it will also include conscious extraterrestrials, even though we
do not have even indirect evidence for them that is so nearly direct as
our (inevitably still indirect) evidence for other conscious beings on
earth.  We can further theorize that if the universe is so large that
there never will be any contact between its distant parts and our part,
there still might be other conscious beings not in causal contact with
us, so that we never could communicate with them to get, even in
principle, the indirect evidence of the same qualitative nature that we
have for other humans here on earth with us.

A next step might be to postulate conscious beings and experiences in
other universes totally disconnected from ours, so that even if one could
imagine traveling faster than the speed of light, there would simply be
no way to get there from here; the two parts would be in totally
disconnected spacetimes.  A similar situation would occur for putative
conscious experiences in other branches of an Everett `many-worlds'
wavefunction or quantum state.  From accepting the existence of such
disconnected observers, it hardly seems like an excessive additional step
to imagine observers in universes or parts of the multiverse with
different constants of physics.  One might even imagine observers in
entirely different universes, not related to ours in the way an entire
multiverse might be related by having one single over-arching set of
natural laws.

So in this sense, the idea of a multiverse seems to be rather a natural
extension of our usual ideas of accepting a reality beyond one's
immediate conscious perception, which is all the experience for which
one has direct access.  All the rest of one's knowledge is purely
theoretical, though one's brain (assuming that one's brain exists and
not just the logical minimum of one's immediate conscious awareness as a
single disembodied entity with absolutely nothing else) is apparently
constructed to bring this knowledge into one's awareness without one's
having to be consciously aware of the details of {\it why} one seems to
be aware of the existence of other conscious beings.

Despite the naturalness of the progression of ideas that leads to
multiverse theories, there are various objections to it.  However, none
of the objections seems to me to be convincing, as there are highly
plausible rebuttals to the objections.

\section{Objections to Multiverse Ideas}

\hspace{.20in} A scientific objection to a multiverse theory might be
that the multiverse (beyond our observed part, which is within one
single universe) is not observable or testable.  But if one had precise
theories for single universes and for multiverses that gave the
distributions of different conditions, one could make statistical tests
of our observations (likely or unlikely in each distribution). 
Unfortunately, no such realistic theory exists yet for either a single
universe or a multiverse, so I would agree that at present we simply do
not have any good theories for either to test.

Another objection is that a multiverse is not a clear consequence of any
existing theory.  Although it is beginning to appear to be a consequence
of string/M theory, that is not yet certain, which is why there can be
theorists like David Gross who are still holding out hope that string/M
theory might turn out to be a single-universe theory after all, possibly
enabling theorists (if they could perform the relevant calculations) to
fulfill their wildest dreams of being able to calculate the constants of
physics uniquely from some simple principles.  One first needs to make
string/M theory into a precise theory and calculate its consequences,
whether single universe or multiverse.  And if that theory gives
predictions that do not give a good statistical fit to observations, one
needs to find a better theory that does.

A philosophical objection to a multiverse theory is that it is
extravagant to assume unfathomable numbers of unobservable universes. 
This is a variant upon the psychological gut reaction that surely a
multiverse would be more complex than a single universe, and hence
should be assigned a lower {\it a priori} probability.  But this is not
necessarily so, as I have explained above.  The whole can be simpler
than its parts, as the set of all integers is quite simple, certainly
simpler than nearly all the (arbitrarily large) individual integers that
form its parts.  The mathematical example above of the simple set of
pili-increasers also shows how even a very large finite set can be much
simpler than almost all of its individual members.

As a further rebuttal of the accusation of extravagance, a theist can say
that since God can do anything that is logically possible and that fits
with His nature and purposes, then there is apparently no difficulty for
Him to create as many universes as He pleases.  He might prefer elegance
in the principles by which He creates a vast multiverse over paucity of
universes, i.e., economy of principles rather than economy of materials.

Another philosophical objection to multiverses is that they can be used
to explain anything, and thereby explain nothing.  I would strongly agree
for this criticism of multiverse theories that are too vague or diffuse,
which do not sufficiently restrict the measure on the set of observations
to favor ordered ones such as what we observe.  There is a genuine need
for a multiverse theory not to spread out the probability measure for
observations so thinly that it makes our observation too improbable.  So
this objection would be a valid objection to vast classes of possible
multiverse theories, but I do not see that it is an objection in
principle against a good multiverse theory.  Certainly not just any
multiverse theory is acceptable, and even if simple single-universe
theories do not work for explaining our observations, it will no doubt be
quite a challenge to find a good multiverse theory that does succeed.

Most of the objections I have raised and attempted to answer so far would
apply both to theistic and nontheistic scientists.  However, if one is a
theist, one might imagine that there are additional objections to
multiverse theories, just as some theists had additional objections to
Darwin's theory of evolution beyond the scientific objections that were
also raised when that theory had much less support.

For example, a theist might feel that a multiverse theory would undercut
the fine-tuning argument for the existence of God.  I shall not deny
that it would undercut the argument at the level of the constants of
physics (though I think there would still be such a design argument from
the general apparently elegant structure of the full laws of nature once
they are known).  However, the loss of one argument does not mean that
its conclusion is necessarily false.

I personally think it might be a theological mistake to look for fine
tuning as a sign of the existence of God.  I am reminded of the exchange
between Jesus and the religious authorities recorded in the Gospel of
Matthew 12:38-41:  ``Then some of the scribes and Pharisees answered,
saying, `Teacher, we want to see a sign from You.'  But He answered and
said to them, `An evil and adulterous generation seeks after a sign, and
no sign will be given to it except the sign of the prophet Jonah.  For as
Jonah was three days and three nights in the belly of the great fish, so
will the Son of Man be three days and three nights in the heart of the
earth.  The men of Nineveh will rise up in the judgment with this
generation and condemn it, because they repented at the preaching of
Jonah; and indeed a greater than Jonah is here.'"  In other words, I
regard the death and resurrection of Jesus as the sign given to us that
He is indeed the Son of God and Savior He claimed to be, rather than
needing signs from fine tuning.

Another theistic objection might be that with a multiverse explanation of
the constants of physics, there is nothing left for God to design.  But
God could well have designed the entire multiverse, choosing elegant laws
of nature by which to create the entire thing.  In any case, whatever the
design is, whether a logically rigid requirement, a simple free choice
God made, or a complex free choice God made, theists would ascribe to God
the task of creating the entire universe or multiverse according to this
design.

A third more specifically Christian objection might be that if the
multiverse (or even just our single part of the universe) is large enough
for other civilizations to have sinned and needed Christ to come redeem
them by something similar to His death on the Cross here on earth for our
sins, then His death may not sound so unique as the Bible says in Romans
6:10:  ``For the death that He died, He died to sin once for all; but the
life that He lives, He lives to God.''  But the Bible was written for us
humans here on earth, so it seems unreasonable to require it to describe
what God may or may not do with other creatures He may have created
elsewhere.  We could just interpret the Bible to mean that Christ's death
here on earth is unique for our human civilization.


\section{Conclusions}

\hspace{.20in} In conclusion, multiverses are serious ideas of present
science, though certainly not yet proven.  They can potentially explain
fine-tuned constants of physics but are not an automatic panacea for
solving all problems; only certain multiverse theories, of which we have
none yet in complete form, would be successful in explaining our
observations.  Though multiverses should not be accepted uncritically as
scientific explanations, I would argue that theists have no more reason
to oppose them then they had to oppose Darwinian evolution when it was
first proposed.

God might indeed so love the multiverse.

\section*{Acknowledgments}

\hspace{.20in} I am indebted to discussions with Andreas Albrecht, Denis
Alexander, Stephen Barr, John Barrow, Nick Bostrom, Raphael Bousso,
Andrew Briggs, Peter Bussey, Bernard Carr, Sean Carroll, Brandon Carter,
Kelly James Clark, Gerald Cleaver, Francis Collins, Robin Collins, Gary
Colwell, William Lane Craig, Paul Davies, Richard Dawkins, William
Dembski, David Deutsch, Michael Douglas, George Ellis, Debra Fisher,
Charles Don Geilker, Gary Gibbons, J.~Richard Gott, Thomas Greenlee, Alan
Guth, James Hartle, Stephen Hawking, Rodney Holder, Richard Hudson, Chris
Isham, Renata Kallosh, Denis Lamoureux, John Leslie, Andrei Linde, Robert
Mann, Don Marolf, Greg Martin, Alister McGrath, Gerard Nienhuis, Cathy
Page, Gary Patterson, Alvin Plantinga, Chris Polachic, John Polkinghorne,
Martin Rees, Hugh Ross, Peter Sarnak, Henry F.~Schaefer III, Paul
Shellard, James Sinclair, Lee Smolin, Mark Srednicki, Mel Stewart,
Jonathan Strand, Leonard Susskind, Richard Swinburne, Max Tegmark, Donald
Turner, Neil Turok, Bill Unruh, Alex Vilenkin, Steven Weinberg, Robert
White, and others whose names I don't recall right now, on various
aspects of this general issue, though the opinions expressed are my own. 
My scientific research on the multiverse is supported in part by the
Natural Sciences and Research Council of Canada.


\begin{thebibliography}{99}

\baselineskip 16.4 pt

\bibitem{Bible} This and all other Scripture taken from the New King
James Version.  Copyright \copyright 1982 by Thomas Nelson, Inc.  Used
by permission.  All rights reserved.

\bibitem{Warfield} Mark Noll and David Livingstone, eds., {\em
B.~B.~Warfield:  Evolution, Science and Scripture} (Baker Books, Grand
Rapids, Michigan, USA, 2000).

\bibitem{FCollins} Francis S.~Collins, {\em The Language of God:  A
Scientist Presents Evidence for Belief} (Free Press, New York, 2006).

\bibitem{Ross83} Hugh Ross, {\em Genesis One: A Scientific Perspective}
(Reasons to Believe, Pasadena, California, 1983).

\bibitem{Ross88} Hugh Ross, {\em Design and the Anthropic Principle}
(Reasons to Believe, Pasadena, California, 1983).

\bibitem{Ross89} Hugh Ross, {\em The Fingerprint of God} (Promise,
Orange, California, 1989).

\bibitem{Swinburne90} Richard Swinburne, ``Argument from the Fine-Tuning
of the Universe,'' in John Leslie (ed.), {\em Physical Cosmology and
Philosophy} (Macmillan, New York, 1990), pp. 154-73; and in John Leslie
(ed.), {\em Modern Cosmology and Philosophy} (Prometheus, Amherst, New
York, 1998), pp. 160-179.

\bibitem{Swinburne91} Richard Swinburne, {\em The Existence of God}
(Clarendon, 1991).

\bibitem{Ross01} Hugh Ross, {\em The Creator and the Cosmos}, 3rd ed.
(NavPress, Colorado Springs, Colorado, 2001).

\bibitem{Manson} Neil A.~Manson, ed., {\em God and Design:  The
Teleological Argument and Modern Science} (Routledge, London, 2003).

\bibitem{Holder} Rodney D.~Holder, {\em God, the Multiverse, and
Everything:  Modern Cosmology and the Argument from Design} (Ashgate,
Burlington, Vermont, USA, 2004).

\bibitem{Mann} Robert B.~Mann, ``Inconstant Multiverse,'' {\it
Perspectives on Science \& Christian Faith} {\bf 57}, 302-310 (2005).

\bibitem{Leslie1} John Leslie, {\em Universes} (Routledge, London and
New York, 1989).

\bibitem{Leslie2} John Leslie, {\em Infinite Minds} (Clarendon Press,
Oxford, 2001).

\bibitem{Barr} Stephen Barr, {\em Modern Physics and Ancient Faith}
(University of Notre Dame Press, Notre Dame, Indiana, 2003).

\bibitem{Collins01} Robin Collins, ``Design and the Many-Worlds
Hypothesis,'' in William Lane Craig, ed., {\em Philosophy of Religion: A
Reader and Guide} (Rutgers University Press, New Brunswick, New Jersey,
2002), pp. 130-148.

\bibitem{Collins05} Robin Collins, ``Design and the Designer:  New
Concepts, New Challenges,'' in Charles S.~Harper, Jr., ed., {\em
Spiritual Information: 100 Perspectives on Science and Religion}
(Templeton Foundation Press, West Conshohocken, Pennsylvania, 2005), pp.
161-167.

\bibitem{Collins07} Robin Collins, ``The Multiverse Hypothesis:  a
Theistic Perspective,'' in Bernard Carr, ed., {\em Universe or
Multiverse?} (Cambridge University Press, Cambridge, 2007), pp. 459-480.

\bibitem{Cleaver03} Gerald Cleaver, ``String/M-Theory Cosmology: God's
Blueprint for the Universe,'' presented at ASA 2003, National Faculty
Leadership Conference 2004, and International Institute for Christian
Studies 2005.

\bibitem{Cleaver06} Gerald Cleaver, ``Before the Big Bang: String
Theory, God, \& the Origin of the Universe,'' presented at Metanexus
2006,
$<$www.metanexus.net/conferences/pdf/conference2006/Cleaver.pdf$>$.

\bibitem{Strand-Kraay} I thank Jonathan Strand (private communication)
for drawing my attention to Klaas Kraay, ``Theism and the Multiverse''
and ``Theism and Modal Collapse'' at
$<$http://www.ryerson.ca/$\sim$kraay/$>$, which for more purely
philosophical reasons support the view that God created a multiverse and
cite many other theists who have also proposed multiverses, often as a
solution for the problem of evil, as I discuss in my companion paper at
$<$http://arxiv.org/abs/0801.0247$>$.

\bibitem{Carter74} Brandon Carter, ``Large Number Coincidences and the
Anthropic Principle in Cosmology,'' in M.~S.~Longair (ed.), {\em
Confrontation of Cosmological Theory with Observational Data} (Riedel,
Dordrecht, 1974), pp. 291-298; reprinted in John Leslie (ed.), {\em
Physical Cosmology and Philosophy} (Macmillan, New York, 1990),
pp. 125-133.

\bibitem{Carr-Rees} Bernard J.~Carr and Martin J.~Rees, ``The Anthropic
Principle and the Structure of the Physical World,'' {\it Nature} {\bf
278}, 605-612 (1979). 

\bibitem{Davies82} Paul Davies, {\em The Accidental Universe}, Cambridge
University Press, Cambridge, 1982).

\bibitem{Carter83} Brandon Carter, ``The Anthropic Principle and its
Implications for Biological Evolution,'' {\it Philosophical Transactions
of the Royal Society of London} {\bf A310}, 347-363 (1983).

\bibitem{BarTip} John D.~Barrow and Frank J.~Tipler, {\em The Anthropic
Cosmological Principle}, (Clarendon Press, Oxford, 1986).

\bibitem{Davies92} Paul Davies, {\em The Mind of God} (Simon \&
Schuster, New York, 1992).

\bibitem{Rees} Martin Rees, {\em Just Six Numbers} (Basic Books, New
York, 2000).

\bibitem{Barrow} John D.~Barrow, {\em The Constants of Nature} (Pantheon
Books, New York, 2002).

\bibitem{Bostrom} Nick Bostrom, {\em Anthropic Bias: Observation
Selection Effects in Science and Philosophy}, (Routledge, New York and
London, 2002).

\bibitem{Susskind} Leonard Susskind, {\em The Cosmic Landscape:  String
Theory and the Illusion of Intelligent Design} (Little, Brown and
Company, New York, 2006).

\bibitem{Davies07} Paul Davies, {\em Cosmic Jackpot:  Why Our Universe
Is Just Right for Life} (Houghton Mifflin, Boston, 2007).

\bibitem{Carr} Bernard Carr, ed., {\em Universe or Multiverse?}
(Cambridge University Press, Cambridge, 2007).

\bibitem{Carter07a} Brandon Carter, ``Objective and Subjective Time in
Anthropic Reasoning,'' arXiv:0708.2367
$<$http://arxiv.org/abs/0708.2367$>$.

\bibitem{Carter07b} Brandon Carter, ``The Significance of Numerical
Coincidences in Nature,'' arXiv:0710.3543
$<$http://arxiv.org/abs/0710.3543v1$>$.

\bibitem{Page-Carr} Don N.~Page, ``Predictions and Tests of Multiverse
Theories,'' in Bernard Carr, ed., {\em Universe or Multiverse?}
(Cambridge University Press, Cambridge, 2007), pp. 411-429,
hep-th/0610101 $<$http://arxiv.org/abs/hep-th/0610101$>$.

\bibitem{Lewis} David K.~Lewis, {\em On the Plurality of Worlds}
(Blackwell, Cambridge, 1986).

\bibitem{Tegmark} Max Tegmark, ``Is `The Theory of Everything' Merely
the Ultimate Ensemble Theory?'' {\it Annals of Physics} {\bf 270}, 1-51
(1998), gr-qc/9704009 $<$http://arxiv.org/abs/gr-qc/9704009$>$.

\bibitem{Teg07} Max Tegmark, ``The Mathematical Universe,'' to be
published in {\it Foundations of Physics}, arXiv:0704.0646
$<$http://arxiv.org/abs/0704.0646$>$.

\bibitem{many-worlds} Bryce DeWitt and R.~Neill Graham, eds., {\em The
Many-Worlds Interpretation of Quantum Mechanics} (Princeton University
Press, Princeton, New Jersey, USA, 1973).

\bibitem{Linde} Andrei D.~Linde, {\em Particle Physics and Inflationary
Cosmology} (Harwood, Chur, Switzerland, 1990).

\bibitem{Guth} Alan Guth, {\em The Inflationary Universe: The Quest for
a New Theory of Cosmic Origins} (Addison-Wesley, Reading, Massachusetts,
USA, 1997).

\bibitem{Vilenkin} Alex Vilenkin, {\em Many Worlds in One: The Search
for Other Universes} (Hill \& Wang, New York, USA, 2006).

\bibitem{DouKac} Michael R.~Douglas and Shamit Kachru, ``Flux
Compactification,'' {\it Reviews of Modern Physics} {\bf 79}, 733-796
(2007), hep-th/0610102 $<$http://arxiv.org/abs/hep-th/0610102$>$.

\bibitem{Swinburne} Richard Swinburne, ed. {\em Bayes's Theorem: 
Proceedings of the British Academy 113} (Oxford University Press,
Oxford, 2002).

\bibitem{Deutsch} David Deutsch, {\em The Fabric of Reality} (Penguin
Books, London 1998).

\bibitem{Deutsch1} David Deutsch, private communication (Jan.~22, 2007).

\bibitem{Woit} Peter Woit, {\em Not Even Wrong:  The Failure of String
theory and the Search for Unity in Physical Law} (Basic Books, New York,
2006).

\bibitem{Littlewood} John E.~Littlewood, ``Sur la Distribution des
Nombres Premiers,'' {\it Comptes Rendus de l'Acad\'{e}mie des Sciences,
Paris} {\bf 158}, 1869-1872 (1914).

\bibitem{Bays-Hudson} Carter Bays and Richard H.~Hudson, ``A New Bound
for the Smallest $x$ with $\pi(x) > \mathrm{li}(x)$,'' {\it Mathematics
of Computation} {\bf 69}, 1285-1296 (2000).

\end{thebibliography}
\end{document}